\documentclass[sigconf,natbib=true,anonymous=false]{acmart}
\usepackage{enumitem}
\usepackage{bm}
\usepackage{multicol}
\usepackage{multirow}
\usepackage{subfigure}
\usepackage{algorithmic}
\usepackage{algorithm}
\usepackage{hyperref}

\AtBeginDocument{%
  \providecommand\BibTeX{{%
    \normalfont B\kern-0.5em{\scshape i\kern-0.25em b}\kern-0.8em\TeX}}}

\setcopyright{acmlicensed}
\copyrightyear{2018}
\acmYear{2018}
\acmDOI{XXXXXXX.XXXXXXX}

\acmConference[Conference acronym 'XX]{Make sure to enter the correct
  conference title from your rights confirmation emai}{June 03--05,
  2018}{Woodstock, NY}
\acmISBN{978-1-4503-XXXX-X/18/06}

\begin{document}

\title{Multi-Grained Preference Enhanced Transformer for Multi-Behavior Sequential Recommendation}


\author{Chuan He}
\affiliation{%
  \institution{Zhejiang University of Technology}
  \city{Hangzhou}
  \country{China}
  }
\email{hechuan@zjut.edu.cn}

\author{Yongchao Liu}
\affiliation{%
  \institution{Ant Group}
    \city{Hangzhou}
  \country{China}
  }
\email{yongchao.ly@antgroup.com}

\author{Qiang Li}
\affiliation{%
  \institution{Zhejiang University of Technology}
    \city{Hangzhou}
  \country{China}
  }
\email{qiangli@zjut.edu.cn}

\author{Weiqiang Wang}
\affiliation{%
  \institution{Ant Group}
    \city{Hangzhou}
  \country{China}
  }\email{weiqiang.wwq@antgroup.com}

\author{Xing Fu}
\affiliation{%
  \institution{Ant Group}
    \city{Hangzhou}
  \country{China}
  }
\email{zicai.fx@antgroup.com}

\author{Xinyi Fu}
\affiliation{%
  \institution{Ant Group}
    \city{Hangzhou}
  \country{China}
  }
\email{fxy122992@antgroup.com}

\author{Chuntao Hong}
\affiliation{%
  \institution{Ant Group}
    \city{Hangzhou}
  \country{China}
  }
\email{chuntao.hct@antgroup.com}

\author{Xinwei Yao$^{\ast}$}
\thanks{*Corresponding author}
\affiliation{%
  \institution{Zhejiang University of Technology}
    \city{Hangzhou}
  \country{China}
  }
\email{xwyao@zjut.edu.cn}

\begin{abstract}
  Sequential recommendation (SR) aims to predict the next purchasing item according to users' dynamic preference learned from their historical user-item interactions. To improve the performance of recommendation, learning dynamic heterogeneous cross-type behavior dependencies is indispensable for recommender system. However, there still exists some challenges in Multi-Behavior Sequential Recommendation (MBSR). On the one hand, existing methods only model heterogeneous multi-behavior dependencies at behavior-level or item-level, and modeling interaction-level dependencies is still a challenge. On the other hand, the dynamic multi-grained behavior-aware preference is hard to capture in interaction sequences, which reflects interaction-aware sequential pattern. To tackle these challenges, we propose a \underline{M}ulti-\underline{G}rained \underline{P}reference enhanced \underline{T}ransformer framework (M-GPT). First, M-GPT constructs an interaction-level graph of historical cross-typed interactions in a sequence. Then graph convolution is performed to derive interaction-level multi-behavior dependency representation repeatedly, in which the complex correlation between historical cross-typed interactions at specific orders can be well learned. Secondly, a novel multifaceted transformer architecture equipped with multi-grained user preference extraction is proposed to encode the interaction-aware sequential pattern enhanced by capturing temporal behavior-aware multi-grained preference . Experiments on the real-world datasets indicate that our method M-GPT consistently outperforms various state-of-the-art recommendation methods. Our code is available at: \href{https://anonymous.4open.science/r/MGPT-DF31}{https://anonymous.4open.science/r/MGPT-DF31}.
\end{abstract}





\maketitle

\section{Introduction}
  Recommendation system has been widely utilized on online platform (e.g., E-commerce sites\cite{DIN}, social media platforms\cite{socialmedia}) to alleviate the information overload and meet users’ diverse interest. Recently, Sequential Recommendation (SR)\cite{sequentialsurvey} has become increasingly essential in recommender system due to its capability of capturing time-varying user preference in regard to historical user-item interactions. 
  \begin{figure}[h]
  \centering
  \includegraphics[width=\linewidth]{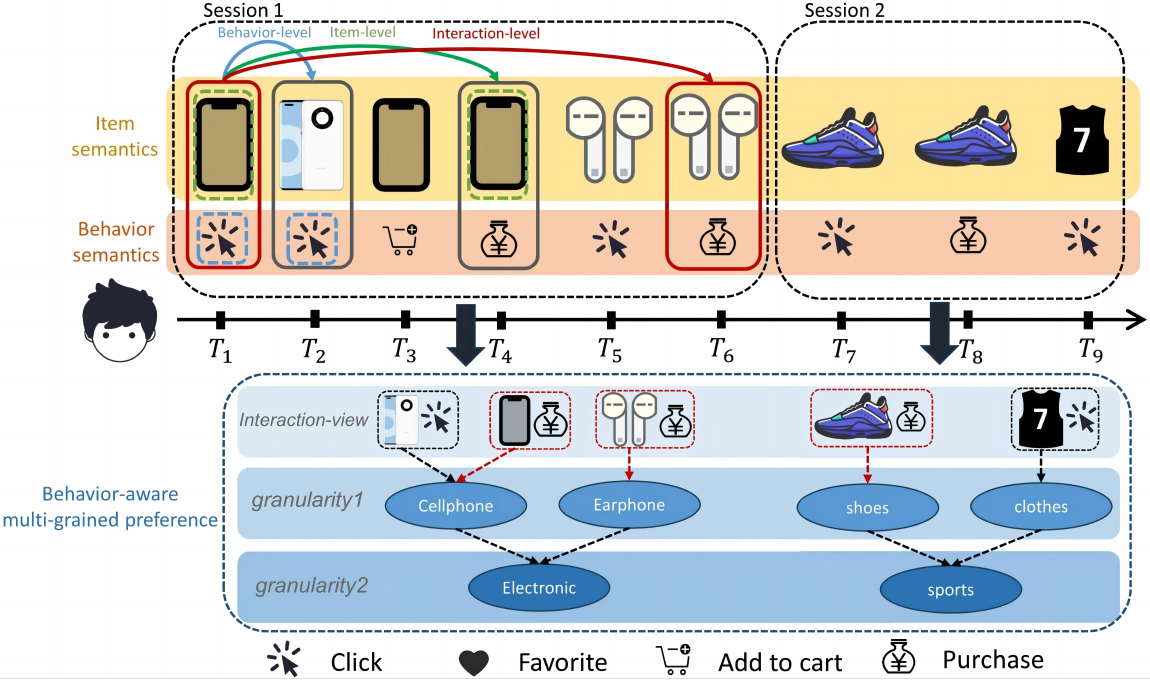}
  \caption{motivation of our work}
  \Description{}
  \vspace{-0.3cm}
\end{figure}
  With the development of deep learning techniques in recent years, a lot of neural network techniques has been applied in solving the sequential recommendation problem. For instance, recurrent neural network-based models aim to learn the sequential pattern within the users’ historical item sequence (e.g., GRU4Rec\cite{GRU4Rec}). Moreover, inspired by the transformer framework, some models propose to leverage self-attention mechanism to encode the item-item dependencies (e.g., BERT4Rec\cite{BERT4Rec}). In addition, graph neural network-based models utilize message passing to learn item transition over the constructed user-item or item-item graph (e.g., SURGE\cite{SURGE}). Nevertheless, the above models only consider a single type of user-item interactions which ignore the heterogeneous cross-type behavior inter-dependency. Interactions on an E-commerce platform encompass a variety of behaviors, such as clicking, adding to favorites, adding to cart, and making a purchase. This multi-behavioral nature provides two key advantages. Firstly, different behaviors, like clicking and making a purchase, indicate distinct user intentions. Therefore, analyzing the multi-behavior interaction data offers an opportunity to capture the nuanced and evolving interests of users. Secondly, the data for the target behavior (e.g., purchases on E-commerce platforms, which are typically of utmost concern) is often sparse, leading to significant cold-start issues when modeling target behavior data independently. A few pioneering research has shifted its focus to multi-behavior sequential recommendation (MBSR) problem. Different from the single-interaction data, multi-behavioral data provides various views of user preference. Some models propose to capture the correlation among behavior-specific sub-sequence (e.g., DMT\cite{DMT}, GNNH\cite{yu2022graph}). Meanwhile, the other research explores the behavior-aware item transitions through injecting the behavior interactions (e.g., MBGCN\cite{MBGCN}, MB-GMN\cite{MB-GMN}, MBHT\cite{MBHT}, MB-STR\cite{MB-STR}).

  Although these previous works are successful in modeling the sequential patterns in behavioral view, there still exist some challenges in MBSR problem:

\begin{itemize}[labelsep=5pt, leftmargin=10pt]
       \item \textbf{CH1. Learning Interaction-Level Dependencies}.In multi-behavior recommendation scenario, a historical interaction consist of item-specific semantic and behavior-specific semantic (e.g., click, favorite, add to cart and purchase). Items with varying behaviors interacting will give rise to intricate multi-behavior dependencies. Some prior approaches (e.g., MB-GMN\cite{MB-GMN}, MB-GCN\cite{MBGCN}, and DMT\cite{DMT}) involve initially aggregating items within each behavior to obtain a cohesive representation, followed by modeling dependencies across all behaviors using attention or weighted summation operations. These approaches model the multi-behavior dependency between interactions with same behavior type, which we called \textbf{behavior-level} dependency (e.g., blue arrow). Recently, MB-STR\cite{MB-STR} proposed multi-behavior multi-head self-attention to model multi-behavior dependency between interactions with same item, which we called \textbf{item-level} dependency (e.g., green arrow). Nevertheless, multi-behavior dependency between interactions with inconsistent behavior types and items is significant as well, which we defined as \textbf{interaction-level} dependency (e.g., red arrow). For instance, purchasing cell phone increases the probability of click on earphone. And there are rarely methods to model it. Thus, \textit{how to model multi-behavior dependencies at \textbf{interaction-level}} is still a challenge for multi-behavior recommendation.
       \item \textbf{CH2. Learning Dynamic Behavior-Aware Multi-Grained Preference}.Sequential information is significant to multi-behavior sequential recommendation. A long-term interaction sequence can be divided into several sessions according to users’ dynamic multi-grained preference. As depicted in Figure 1, the interaction sequence of the boy consists of two main sessions. The first session including cell phones and earphones reflects the intention for electronic products. The second session including basketball shoes and basketball jerseys reveals the interest for sports. These two varying sessions show the \textbf{dynamic user preference}. From a deep perspective of a single session, when we only focus on one interaction, such as purchasing a pair of basketball shoes, we may only think that he is interested in collecting sneakers. However, when we combine other interactions, such as purchasing basketball jerseys, we will find that he is actually interested in playing basketball as a sport, which reveals the \textbf{multi-grained preference}. Meanwhile, we can distinguish the intensity of interest in varying items by \textbf{different typed behaviors}. Some previous works (e.g., MB-STR\cite{MB-STR}, MBHT\cite{MBHT}) proposed methods to model the sequential information ignoring the dynamic behavior-aware multi-grained preference. MISSL\cite{MISSL} emphasizes on capturing multi-typed interests. Hence, \textit{how to model multi-behavior sequential pattern involving \textbf{dynamic behavior-aware multi-grained preference}} is still a challenge for multi-behavior sequential recommendation.
\end{itemize}
To solve aforementioned issues, in this paper, we propose \underline{M}ulti-\underline{G}rained \underline{P}reference enhanced \underline{T}ransformer (M-GPT) to improve the recommendation performance. To achieve this goal, M-GPT is designed with two core components, e.g., interaction-aware dependency extractor and multifaceted sequential pattern generator. The interaction-level dependency extractor is developed to model the personalized interaction-level multi-behavior dependency from low-order to high-order(Ch1). In this component, we first construct a learnable graph structure according to item-behavior interactions in user historical sequence. Specifically, the incidence matrix of our graph is calculated by the inner product of item- and behavior-level dependency representations, which learns multi-behavior dependency at interaction-level. Then graph convolution is utilized iteratively to model interaction-level dependency in various orders. The multifaceted sequential pattern generator aims to capture the sequential interaction pattern enhanced by extracting behavior-aware multi-grained preference in different time scales (Ch2). We first perform the transformer layer with linear attention to model the sequential pattern more efficiently. Moreover, the interaction sequences are divided into several sessions by different time scales. Then the multi-grained self-attention is performed to capture session-specific multi-behavior preference based on multi-grained user intent in each session. To aggregate the multi-grained preference captured in each session, an aggregator is performed to fuse session-based multi-grained preference representations with temporal effects into our sequential interaction pattern. To summarize, the contributions in this paper are listed as follows,
\begin{itemize}[labelsep=5pt, leftmargin=10pt]
    \item We highlight the existing challenges in multi-behavior sequential recommendation, including modelling interaction-level multi-behavior dependencies and dynamic behavior-aware multi-
    grained preference.
    \item We propose a novel framework M-GPT for multi-behavior sequential recommendation. Two core components are Interaction-Level Dependency Extractor(IDE) and Multifaceted Sequential Pattern Generator(MSPG). Interaction-level multi-behavior dependency is modelled by IDE through a specially designed interaction-aware graph. Moreover, multi-behavior sequential pattern is learned by MSPG enhanced with modelling dynamic behavior-aware multi-grained preference.
    \item We perform extensive experiments on three public datasets, which validate the superiority of our proposed M-GPT compared with some state-of-art recommender systems. More meticulous experiments further show the benefits of our model.
\end{itemize}

\section{Related Work}
\subsection{Sequential Recommendation}
Sequential Recommendation (SR) is designed to capture the evolution of user preferences by modeling sequences. Initial approaches typically modeled sequential dependencies based on the Markov Chain assumption [24]. However, with the progression of deep learning, models based on Recurrent Neural Networks (RNN) \cite{PARSRec, GRU4Rec, DIEN}, Convolutional Neural Networks (CNN) \cite{Caser}, Graph Neural Networks (GNN) \cite{SURGE, SR-GNN, GCSAN, HyperRec, lightgcn, huang2021graph, mlkg, NI-CTR, multiviewgraph, graphmasked, MGNM}, and attention mechanisms \cite{kr-mbsr, LinRec, SASRec, TiSASRec, BERT4Rec, DDIN, DIN} have been utilized to uncover dynamic user interests hidden within behavior sequences. More recently, models based on contrastive learning\cite{dualconstrastive, dualpastfurture, qiu2021memory, wei2022contrastive} have been introduced to extract significant user patterns by generating self-supervision signals. While these methods enhance the performance of sequential recommendation, they tend to have limited predictive power when dealing with short single-behavior sequences.
\subsection{Multi-Behavior Recommendation}
Recently, multi-behavior recommender systems have been developed to model the heterogeneity of user-item relations \cite{NMTR, MBGCN, MB-GMN,zhang2020multiplex,kr-mbsr}. For instance, NMTR\cite{NMTR} is a multi-task recommendation framework that establishes predefined cascading relationships between behaviors. Inspired by the capabilities of Graph Neural Networks (GNNs), models such as MBGCN\cite{MBGCN}, MBGMN \cite{MB-GMN}, and MGNN \cite{zhang2020multiplex} have been developed. These models utilize graph-structured message passing over the generated multi-relational user-item interaction graphs. However, these approaches do not take into account the time-evolving multi-behavior user preference. Some existing work \cite{MB-STR, MBHT}separates the modeling phases of item and behavior sequences. These methods incorporate behavior patterns as auxiliary information by adding behavior types into the input or modeling behavior sequences independently. For instance, BINN \cite{BINN} uses a contextual long short-term memory architecture to model item and behavior sequences. MB-STR \cite{MB-STR} leverages both behavior-specific semantics and multi-behavior sequential heterogeneous dependencies via transformer layers. These methods maintain the integrity of interaction sequences, enabling the exploration of complex multi-behavior sequential patterns.
\section{Methodology}
\subsection{Problem Formulation}
We first describe a typical multi-behavior recommendation scenario. Suppose that we have $\lvert U \rvert$ users $u_i \in U$ and $\lvert V \rvert$ items $v_i \in V$ in our multi-behavior recommender system. In real shopping scenario, there exists various types of user-item interactions like click, favorite, add to cart and purchase. Thus, we formulate the set of behavior $B=\{B_1,B_2,…,B_{\lvert B \rvert}\}$, where $\lvert B \rvert$ is the number of user-item interaction type. Among different types of user-item interactions, purchase is the most important one we care about which is called target behavior and the other behaviors are called auxiliary behaviors. For an individual user $u_i$, his historical user-item interactions compose a personalized multi-behavior interaction sequence, which is defined as $S_u=\{ \left \langle v_1,b_1 \right \rangle,\left \langle v_2,b_2 \right \rangle,…,\left \langle v_{\lvert s_u \rvert} ,b_{\lvert s_u \rvert}  \right \rangle\}$ and $\lvert s_u \rvert$ is the number of interactions in the sequence. Our task is to predict top-\textit{K} items from $V$ that have a higher possibility to be interacted by user $u_i$ under target behavior at the next time step by extracting the latent information in user’s personalized dynamic heterogeneous multi-behavior dependencies.
\begin{figure*}[htbp]
  \centering
  \includegraphics[width=\linewidth]{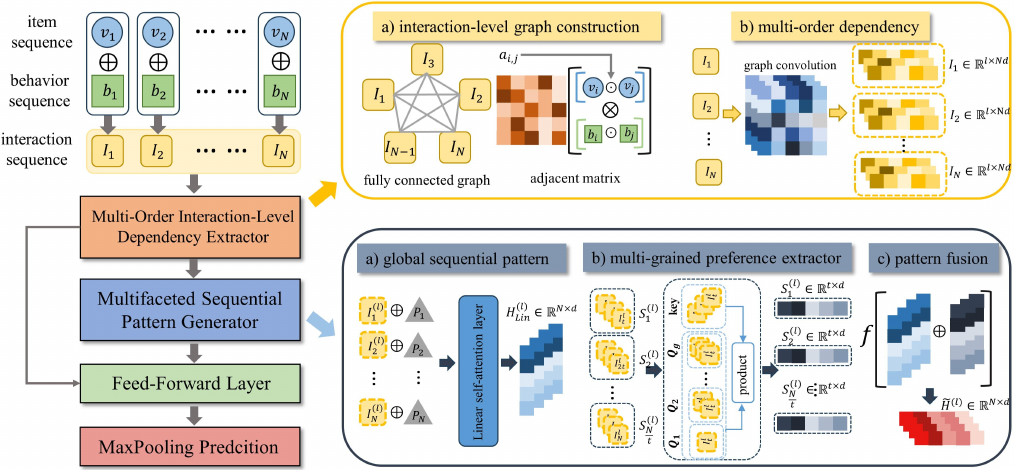}
  \caption{The overview structure of M-GPT}
  \Description{}
\end{figure*}
\subsection{Overview of M-GPT}
In this section, we introduce our proposed M-GPT framework. As depicted in Figure 2, M-GPT consists of two important components: 1) \textbf{interaction-level dependency extractor}, and 2) \textbf{multifaceted sequential pattern generator}. Firstly, to learn the multi-behavior dependencies at interaction-level, we design the interaction-level dependency extractor which is a graph learning paradigm. The graphs are constructed to consider both item-level and behavior-level multi-behavior dependencies. Then graph convolution is performed iteratively to model interaction-level inter-dependencies at different orders. Moreover, to precisely model the sequential patterns of user historical interactions, the multifaceted sequential pattern generator is proposed which follow the transformer-based method. Specifically, we propose multi-grained self-attention mechanism to capture users’ temporal multi-grained preference in different time scales to enhance the expression of sequential patterns. Finally, the model predicts the top-\textit{K} items users intend to purchase through learning the latent information from these two components. The overall learning progress of M-GPT is performed in Appendix \ref{algorithm}.
\subsection{Interaction-Level Dependency Extractor}
In multi-behavior recommendation, multi-behavior dependency consists of item-level dependency and behavior-level dependency. In previous works, these two types of dependencies are modelled in an asynchronous or independent manner, which deteriorates the effectiveness of recommendation. To this end, we propose interaction-level dependency extractor to model multi-behavior dependency at interaction level that models item- and behavior- level dependency in a synchronous and integrated manner. 
\paragraph{3.3.1\quad Interaction-aware Context Embedding Layer}
In multi-
behavior sequential recommendation, user-item interactions consist of item-specific and behavior-specific semantics. To extract the interaction-level dependency efficiently, we design the interaction-aware context embedding layer to jointly encode the item-level and behavior-level contextual information. To this end, we define the interaction-aware latent representation $x_i \in \mathbb{R}^d$ of a given user-item interaction as following:
\begin{equation}
    h_i = e_i \oplus b_i,
\end{equation}
\begin{equation}
    M_{s_u} = \{h_0, h_1,..., h_{{\lvert s_u \rvert}-1}\}
\end{equation}
where $e_i \in \mathbb{R}^d$ represents the item embedding of $v_i$. $b_i \in \mathbb{R}^d$ is the representation of behavior embedding according to the behavior type user $u_i$ interacts with item $v_i$. $M_{s_u}$ is a set of interaction representation within the historical sequence of user $u_i$.
\paragraph{3.3.2\quad Interaction-Level Graph Construction}
Given the historical interaction sequence $M_{s_u} = \{h_0, h_1,..., h_{{\lvert s_u \rvert}-1}\}$ of user $u$, we convert these interactions into a fully connected undirected graph $\mathcal{G}_{s_u}$. To learn \textit{interaction-level} multi-behavior dependency, we introduce the incidence matrix $\mathcal{A} \in \mathbb{R}^{{\lvert s_u \rvert} \times {\lvert s_u \rvert}}$ taking both item-level and behavior-level semantics into consideration. To achieve this goal, we calculate each entry $\mathcal{A}_{i,j} \in \mathbb{R}$ as following,
\begin{equation}
\label{3}
    E_{i,j}=e_i \odot e_j,
\end{equation}
\begin{equation}
\label{4}
    B_{i,j}=b_i \odot b_j,
\end{equation}
\begin{equation}
\label{5}
    \mathcal{A}_{i,j}=E_{i,j} \cdot B_{i,j}
\end{equation}
where $e_i, e_j \in \mathbb{R}^d$ are item semantic embedding representations and $b_i, b_j \in \mathbb{R}^d$ are behavior semantic embedding representations.First, we operate dot product on item-specific pair and behavior-specific pair to get $E_{i,j}, B_{i,j} \in \mathbb{R}^d$ respectively, which learns the multi-behavior dependency from item-level and behavior-level perspectives \textit{synchronously}. To learn item- and behavior-level dependency in an \textit{integrated} manner, an inner product is performed on item-level dependency representation and behavior-level dependency representation. Meanwhile, we add \textit{L}1 regularization on the incidence matrix $\mathcal{A}$ to facilitate the interaction-level dependency learning.
\paragraph{3.3.3\quad Multi-Order Interaction-Level Dependency Learning}
Learning interaction-level dependency in a single order is not desirable\cite{MGNM} due to the diverse complexity of the personalized behavior pattern. Thus, we use graph convolution to gain dependencies representation from low-order to high-order as follows:
\begin{equation}
\label{6}
    H^{(l+1)}=LeakyReLU(\tilde{D}^{-\frac{1}{2}}\tilde{\mathcal{A}}\tilde{D}^{-\frac{1}{2}}H^{(l)}W),
\end{equation}
\begin{equation}
\label{7}
    \tilde{D}^{-\frac{1}{2}}=I+D^{-\frac{1}{2}}\mathcal{A}D^{-\frac{1}{2}},
\end{equation}
\begin{equation}
\label{8}
    H^{(0)}=[{h^{(0)}_0, h^{(0)}_1,..., h^{(0)}_{{\lvert s_u \rvert}-1}}]
\end{equation}
where $H^l$ denotes the interaction-level dependency representations in different order $l \in \{0,...,L\}$. We follow the common practice of graph convolution. identity matrix $I$ is added to  achieve self-loop aggregation, $D$ denotes the degree matrix of $\mathcal{A}$. $W$ is a fixed parameter matrix to achieve the aggregation of high-order neighbors more efficiently and effectively. As for activation function, we use $LeakyReLU$. All the dependency representations at various orders will be utilized to be the input of multifaceted sequential pattern generator.
\subsection{Multifaceted Sequential Pattern Generator}
Existing transformer-based SR methods efficiently model one-sided sequential dependencies of item transitions. Nevertheless, in real world multi-behavior recommendation scenario, there are a lot of factors influencing the generation of multi-behavior sequential patterns. For instance, user $u_i$'s point of interests changes in different time scales. Moreover, users' preference is multi-grained in a single time scale. Thus, it's indispensable for us to design a multifaceted method to extract sequential pattern from the historical interaction sequence. In this section, the details of multifaceted sequential pattern generator will be introduced.
\paragraph{3.4.1\quad Sequential Information Injection}
To encode the sequential pattern of user $u$, we need to inject the sequential information into the interaction-level dependency representation of each historical interactions in sequence $S_u$. Specifically, we set the max length of historical interaction sequence $S_u$ as $N$. if the sequence length is less than $N$, special [padding] tokens are padded to the left as dummy past interactions. Then, we construct a positional embedding matrix $\mathbb{R}^{N \times d}$ to encode the sequential information:
\begin{equation}
    H^{(l)}=[h_0^{(l)} \oplus p_0,...,h_{N-1}^{(l)}  \oplus p_{N-1}]\quad\quad(l=0,1,...)
\end{equation}
where $H^{(l)}$ is the $l$-th order of interaction-level dependency representation, $h_i^{(l)}$ denotes the interaction representation on $i$-th position, $p_i$ denotes the embedding representation of $i$-th position. We inject the sequential information through element-wise add operation. 
\paragraph{3.4.2\quad Global Sequential Pattern Encoding}
Previous works leveraged self-attention layer to encode the global sequential pattern. To alleviate the high computational and memory cost of dot-product for long term sequence, we utilize a linear self-attention layer to encode the global sequential pattern inspired by \cite{LinRec}.

Specifically, to reduce the model complexity to $\bm{O}(N)$, we first calculate $\bm{K}^{\mathrm{T}}\bm{V}$ rather than $\bm{Q}\bm{K}^{\mathrm{T}}$, which reduce the model complexity from $\bm{O}(N^{2}d)$ to $\bm{O}(Nd^2)(N>>d)$. Then, to make our linear attention mechanism equivalent to original dot product attention mechanism, we perform row- and column-wise $L2$ Normalization on $\bm{Q}$(query matrix) and $\bm{K}$(key matrix) respectively. Meanwhile, to ensure a relatively stable learning process, ELU function is performed on the $\bm{Q}$ and $\bm{K}$. The formulation of linear attention mechanism is shown as following:
\begin{equation}
\label{10}
    H_{Lin}^{(l)}=\rho_1(elu(H^{(l)}{W_Q}))(\rho_2(elu(H^{(l)}{W_K}))^{\mathrm{T}}H^{(l)}{W_V})
\end{equation}
\begin{equation}
\label{11}
    \rho_1(elu(H^{(l)}{W_Q})_i)=\frac{elu(H^{(l)}{W_Q})_i}{\sqrt{d}\Vert elu(H^{(l)}{W_Q})_i \Vert_2}
\end{equation}
\begin{equation}
\label{12}
    \rho_2(elu(H^{(l)}{W_K})_j)=\frac{elu(H^{(l)}{W_K})_j}{\sqrt{d}\Vert elu(H^{(l)}{W_K})_j \Vert_2}
\end{equation}
where $\rho_1(\cdot)$ denotes the row-wise $L2$ Normalization for $\forall i \in [N]$ and $\rho_2(\cdot)$ denotes the column-wise $L2$ Normalization for $\forall j \in [d]$, where $elu(H^{(l)}{W_Q})_i$ is $i$-th row of $elu(H^{(l)}{W_Q})$ and $elu(H^{(l)}{W_K})_j$ is $j$-th column of $elu(H^{(l)}{W_K})$. ${W_Q}$, ${W_K}$ and ${W_V}$ are transformation matrices. We perform linear self-attention layer on the whole interaction sequence to learn the global sequential pattern efficiently and effectively.
\paragraph{3.4.3\quad Temporal Multi-Grained Preference Encoding}
The global sequential pattern reflects user's stable long-term preference. Nevertheless, user's short-term preference varies from different time scales which is fluctuated. To model short-term preference, we first divide the interaction sequence into sessions. Given an historical interaction sequence $H^{(l)}=[h_0^{(l)},...,h_{N-1}^{(l)}]$ of user $u$, we divide it into several sessions with different time scales, as following:
\begin{equation}
\label{13}
    H^{(l)}=[S_0^{(l)},S_1^{(l)}...,S_{\frac{N}{t}-1}^{(l)}]
\end{equation}
\begin{equation}
\label{14}
    S_i^{(l)}=[h_{i \times t}^{(l)},h_{i \times t + 1}^{(l)},...,h_{(i+1) \times t -1}^{(l)}]
\end{equation}
where $t$ denotes the number of interactions in a session, $S_i^l$ denotes the $i$-th session and the number of sessions is $\frac{N}{t}$. To construct a hierarchical structure, we can select different values of $t$ to learn various user preference with different \textit{time granularity}.

Moreover, user preference is various at different levels of perspective as we talk about in introduction. Inspired by \cite{Atten-Mixer}, we propose a \textit{multi-grained multi-head self-attention layer} to encode the multi-grained preference in sessions divided by different time scales. First, to create multi-grained user intent, we group the last items with different lengths in a session. Then, we concatenate them within the group to form a raw group representation. At last, linear transformation is performed on these group representations to represent the multi-grained user queries. The detail is shown as following:
\begin{equation}
\begin{split}
\label{15}
    & Q_1=W_{q_1}(h_{t-1}^{(l)}),\\
    & Q_2=W_{q_2}(Concat(h_{t-1}^{(l)}, h_{t-2}^{(l)})), \\
    & ......,\\
    & Q_g=W_{q_g}(Concat(h_{t-1}^{(l)},...,h_{t-g}^{(l)}))\\
\end{split}
\end{equation}
where $h_j^{(l)} \in S_i^{(l)}$ denotes interaction representation in the session of $l$-th order, $W_{q_m} \in \mathbb{R}^{d \times md}$ denotes linear transformation matrices for multi-grained user queries. Generated multi-grained queries representation reflects characteristics of short-term sequence including inherent priority and local invariance.

After generating the multi-grained query representations, we concatenate them into a whole query matrix $Q \in \mathbb{R}^{g \times d}$. Then, multi-head attention layer is performed and the attention weights are calculated as
\begin{equation}
\label{16}
    \alpha_h=softmax(\frac{QW_h^Q(S_i^{(l)}W_h^K)^{\mathrm{T}}}{\sqrt{d}})
\end{equation}
where $S_i^{(l)}\in\mathbb{R}^{t \times d}$ denotes the whole interaction representations. $W_h^Q, W_h^K \in \mathbb{R}^{d \times d}$ are the transformation matrices. $h=1,2,...,H$ denotes the attention head index. We get $\alpha \in \mathbb{R}^{N \times gH}$, the combination of multi-head attention weights, and perform $L_p$ pooling on the weight $\alpha$ to balance the influence of different query granularity.
\begin{equation}
\label{17}
    \tilde{\alpha}_{j,h}=[\sum_{v=0}^{g-1}(\alpha_{j, vH+h})^p]^{\frac{1}{p}}
\end{equation}
\begin{equation}
\label{18}
\tilde{S_i^{(l)}}=Concat(\tilde{\alpha}_{1}S_i^{(l)}W_1^V,...,\tilde{\alpha}_{H}S_i^{(l)}W_H^V)
\end{equation}
where $\tilde{\alpha}_{j,h}$ denotes the output of pooling operator. $\tilde{S_i^{(l)}}$ denotes the multi-grained user preference representation of $i$-th session in $l$-th order dependency representation. After that, we add the average of positional embedding representations in session $S_i^{(l)}$ to encode the sequential information of the session as following:
\begin{equation}
\label{19}
    \tilde{S_i^{(l)}}\oplus=ave(p_{i \times t}\oplus,...,\oplus p_{(i+1) \times t-1})
\end{equation}
Given $\frac{N}{t}$ multi-grained preference representations $\tilde{S_i^{(l)}} \in \mathbb{R}^{t \times d}(i=0,...,\frac{N}{t}-1)$, we get a whole sequence preference representation $\tilde{S_t^l} \in \mathbb{R}^{{N} \times {d}}$ with time scale $t$. In M-GPT, we select two different time scales $t_1$ and $t_2$ to learn multifaceted preference representations.

\paragraph{3.4.4\quad Multifaceted Pattern Fusion}
To fuse our multifaceted pattern representations, we design a fusion layer to aggregate the global pattern embedding $ H_{Lin}^{(l)} \in \mathbb{R}^{N \times d}$ and local pattern embedding $\tilde{S_{t_1}^{(l)}}, \tilde{S_{t_2}^{(l)}} \in \mathbb{R}^{{N} \times {d}}$ enhanced with multi-grained preference as follows:
\begin{equation}
    \tilde{H^{(l)}}=(H_{Lin}^{(l)}\Vert \tilde{S_{t_1}^{(l)}} \Vert \tilde{S_{t_2}^{(l)}})W^d
\end{equation}
where $W^d \in \mathbb{R}^{3N \times N}$ is the projection matrix which transforms $\mathbb{R}^{3N \times d}$ dimensional embedding into $\mathbb{R}^{N \times d}$ dimensional representations. 
\paragraph{3.4.5\quad Multifaceted Transformer Layer}
At last, non-linearity is injected into our multifaceted transformer layer. We also perform residual connection and layer normalization on the output:
\begin{equation}
    \tilde{H^{(l)}}^n=LayerNorm(\tilde{H^{(l)}}^n+\tilde{H^{(l)}}^{n-1})
\end{equation}
\begin{equation}
    \tilde{H^{(l)}}^n=GELU(W_1^n\tilde{H^{(l)}}^{n-1}+b_1^n)W_2^n+b_2^n
\end{equation}
where $W_1^n, W_2^n \in \mathbb{R}^{d \times d_h}$ and $b_1^n,b_2^n \in \mathbb{R}^{d}$ are learnable projection matrices and bias terms. $n$ denotes the $n$-th Multifaceted Transformer Layer.
\subsection{Model Training and Prediction}
\paragraph{3.5.1\quad Prediction}
Given a candidate item $x_t$, we calculate the recommendation score which denotes the probability of $x_t$ being the $i$-th position in interaction sequence as follow:
\begin{equation}
    \tilde{p_{i,t}}=MaxPooling(\tilde{H^{(0)}}_i^{n}e_t, \tilde{H^{(1)}}_i^{n}e_t,...,\tilde{H^{(l)}}_i^{n}e_t)
\end{equation}
where $p_{i,t}$ denotes the probability of item $x_t$ being the $i$-th position in sequence, $\tilde{H^{(l)}}_i^{n}$ denotes the representation embedding of $i$-th position in $l$-th order of interaction-level dependency representations and $e_t$ denotes the item embedding of item $x_t$. In M-GPT, we choose to  perform MaxPooling on various orders of interaction-level dependency rather than SumPooling or using attention mechanism. 
\paragraph{3.5.2\quad Training}
We train proposed M-
GPT by Cloze task\cite{kang2021entangled, BERT4Rec, MBHT}. Specifically, we first mask some interactions whose behavior type is target behavior(e.g. purchase) with mask ratio $\rho$. Then, we replace these masked interactions including their item embedding and behavior embedding with special token [MASK], which follows the training strategy as \cite{MBHT}. For enabling our M-GPT get best performance from low-order to high-order interaction-level dependency, we tend to define a cross-entropy loss for each order. Hence, our loss $L$ is defined as follows:
\begin{equation}
\mathcal{L}_{all}=\sum_{l=1}^{l}\mathcal{L}_l+\theta_{1}\mathcal{L}_1+\theta_{2}\mathcal{L}_2
\end{equation}
\begin{equation}
    \mathcal{L}_l=\frac1{|T|}\sum_{t\in T,i\in I}-\log(\frac{\exp\tilde{H^{(l)}}_i^{n}e_t}{\sum_{j\in V}\exp\tilde{H^{(l)}}_i^{n}e_j})
\end{equation}
where $T$ is the set of ground-truth ids for masked items in each
batch, $I$ is the set of masked positions corresponding to $T$, and $V$
is the item set. $L_1$ denotes the $L_1$ normalization of adjacent matrix $\mathcal{A}$ and $L_2$ denotes the $L_2$ normalization of model parameters. $\theta_{1}$ and $\theta_{2}$ denote the hyperparameters. The time complexity analysis is shown in Appendix \ref{time complexity}.
\section{EXPERIMENT}
In this section, we conduct comprehensive experiment on three real-world datasets to answer the following questions:
\begin{itemize}[labelsep=5pt, leftmargin=10pt]
    \item \textbf{RQ1}: How does our \textbf{M-GPT} perform against various state-of-the-art recommendation methods?
    \item \textbf{RQ2}: If our proposed module(e.g. interaction-level dependency extractor, multifaceted sequential pattern generator) works effectively in \textbf{M-GPT}?
    \item \textbf{RQ3}: How does the performance of \textbf{M-GPT} vary with different values of hyper-parameters?
    \item \textbf{RQ4}: If users' diverse behavior patterns can be well learned by \textbf{M-GPT}?
    \item \textbf{RQ5}: How to prove that \textbf{M-GPT} works in an interpretable way?

The comprehensive result of hyper-parameter analysis is provided in Appendix \ref{hyper}.
\end{itemize}   
\subsection{Experimental Settings}
\paragraph{4.1.1\quad Dataset.}
To evaluate the performance of our proposed M-GPT, we select two datasets from real-world scenarios. i) \textbf{Taobao}. This dataset is collected from Taobao which is one of the largest e-commerce platforms. It contains four types of behaviors, i.e., \textit{page view}, \textit{tag-as-favorite}, \textit{add-to-cart} and \textit{purchase}. ii) \textbf{IJCAI}. IJCAI is released by IJCAI contest 2015 for repeat buyers prediction from an online business-to-consumer e-commerce, which includes four types of behaviors, i.e.,  \textit{page view}, \textit{tag-as-favorite}, \textit{add-to-cart} and \textit{purchase}. iii) \textbf{Retailrocket}. This dataset was collected from an online shopping website called Retailrocket, spanning a period of 4 months including three behaviors, i.e.,  \textit{page view}, \textit{add-to-cart} and \textit{purchase}. We set the target behavior as \textit{purchase} for dataset Taobao, IJCAI and Retail. For a fair comparison, we closely follow the pre-processed datasets with \cite{MBHT}. The detail about three datasets is shown in Table1.
\begin{table}
  \caption{Statistics of the used dataset}
  \small
  \label{tab:freq}
  \begin{tabular}{ccccc}
    \toprule
    Dataset&\#users&\#items&\#interactions&Behavior types\\
    \midrule
    Taobao & 147,894& 99,037& 7,658,926&\{pv,fav,cart,buy\}\\
    IJCAI & 423,423& 874,328&36,222,123&\{pv,fav,cart,buy\}\\
    Retailrocket & 11,649& 36,233&87,822&\{pv,cart,buy\}\\
  \bottomrule
\end{tabular}
\end{table}
\paragraph{4.1.2\quad Evaluation Protocols}
Following the settings in MBHT, we adopt the leave-one-out strategy for performance evaluation. We choose two evaluation metrics,i.e., Hit Rate (HR), Normalized Discounted Cumulative Gain(NDCG) and Mean Reciprocal Rank(MRR), where we set the cut-off of ranked lists as 5 and 10. For all experiments, we select the average performance of five times.
\begin{table*}[htbp]
  \caption{Experimental results on two datasets. The best results are boldfaced and the second-best results are underlined}
  \scriptsize
  \label{tab:freq}
  \begin{tabular}{c|ccccc|ccccc|ccccc}
    \toprule
    \multirow{2}{*}{Model} & \multicolumn{5}{c|}{Taobao} & \multicolumn{5}{c}{IJCAI} & \multicolumn{5}{|c}{Retail} \\
    \cline{2-6} \cline{7-11} \cline{12-16}
    \quad & HR@5 & NDGC@5 & HR@10 & NDGC@10 & MRR & HR@5 & NDGC@5 & HR@10 & NDGC@10 & MRR & HR@5 & NDGC@5 & HR@10 & NDGC@10 & MRR \\
    \midrule
    \hline
    Caser & 0.081& 0.060 & 0.123 & 0.070 & 0.067 & 0.122 & 0.081 & 0.155 & 0.102 & 0.108 & 0.581 & 0.490 & 0.703 & 0.530 & 0.488\\
    GRU4Rec & 0.142 & 0.101 & 0.207 & 0.120 & 0.115 & 0.132 & 0.094 & 0.189 & 0.110 & 0.112 & 0.590 & 0.527 & 0.655 & 0.598 & 0.520\\
    SASRec & 0.150 & 0.108 & 0.203 & 0.122 & 0.120 & 0.137 & 0.101 & 0.189 & 0.119 & 0.116 & 0.620 & 0.597 & 0.641 & 0.601 & 0.590\\
    HPMN & 0.160 & 0.130 & 0.213 & 0.138 & 0.130 & 0.139 & 0.093 & 0.194 & 0.122 & 0.121 & 0.612 & 0.580 & 0.665 & 0.538 & 0.555\\
    BERT4Rec & 0.197 & 0.154 & 0.252 & 0.174 & 0.161 & 0.286 & 0.209 & 0.392 & 0.239 & 0.216 & 0.759 & 0.622 & 0.832 & 0.644 & 0.591\\
    SR-GNN & 0.098 & 0.070 & 0.151 & 0.086 & 0.086 & 0.067 & 0.048 & 0.110 & 0.061 & 0.058 & 0.797 & 0.728 & 0.840 & 0.741 & 0.719\\
    GCSAN & 0.217 & 0.157 & 0.302 & 0.189 & 0.171 & 0.101 & 0.082 & 0.164 & 0.099 & 0.097 & 0.821 & 0.799 & 0.843 & 0.803 & 0.791\\
    HyperRec & 0.141 & 0.131 & 0.217 & 0.133 & 0.126 & 0.129 & 0.100 & 0.224 & 0.137 & 0.121 & 0.813 & 0.854 & 0.785 & 0.772 & 0.767\\
    SURGE & 0.122 & 0.080 & 0.188 & 0.101 & 0.093 & 0.211 & 0.142 & 0.309 & 0.177 & 0.162 & 0.855 & 0.821 & 0.837 & 0.841 & 0.825\\
    \hline
    MB-GCN & 0.185 & 0.103 & 0.309 & 0.143 & 0.149 & 0.218 & 0.145 & 0.335 & 0.182 & 0.177 & 0.830 & 0.689 & 0.798 & 0.701 & 0.688\\
    NMTR & 0.125 & 0.082 & 0.174 & 0.097 & 0.103 & 0.109 & 0.076 & 0.184 & 0.099 & 0.106 & 0.810 & 0.651 & 0.780 & 0.677 & 0.692\\
    MB-GMN & 0.192 & 0.108 & 0.319 & 0.154 & 0.151 & 0.235 & 0.161 & 0.337 & 0.193 & 0.176 & 0.853 & 0.709 & 0.804 & 0.786 & 0.779\\
    \hline
    BERT4Rec-MB & 0.209 & 0.170 & 0.260 & 0.186 & 0.178 & 0.241 & 0.179 & 0.330 & 0.203 & 0.190 & 0.825 & 0.809 & 0.841 & 0.814 & 0.805\\
    NextIP & 0.301 & 0.244 & 0.387 & 0.274 & 0.241 & 0.303 & 0.232 & 0.391 & 0.258 & 0.245 & 0.899 & 0.893 & 0.906 & 0.894 & 0.892\\
    MB-STR & 0.309 & 0.248 & 0.394 & 0.278 & 0.250 & 0.310 & 0.236 & 0.393 & 0.261 & 0.251 & 0.907 & \underline{0.899} & 0.910 & \underline{0.899} & \underline{0.896}\\
    MBHT & 0.320 & 0.253 & 0.402 & 0.280 & 0.259 & 0.306 & 0.238 & 0.390 & 0.265 & 0.246 & \underline{0.908} & 0.897 & \underline{0.912} & 0.898 & 0.895\\
    DyMus+ & 0.287 & 0.198 & 0.313 & 0.223 & 0.230 & 0.289 & 0.217 & 0.384 & 0.247 & 0.228 & 0.862 & 0.855 & 0.883 & 0.878 & 0.879\\
    TGT & 0.289 & 0.202 & 0.312 & 0.226 & 0.232 & 0.288 & 0.219 & 0.381 & 0.250 & 0.230 & 0.883 & 0.867 & 0.901 & 0.892 & 0.887\\
    PBAT & \underline{0.329} & \underline{0.262} & \underline{0.410} & \underline{0.286} & \underline{0.264} & 0.308 & 0.235 & 0.390 & 0.262 & 0.250 & 0.907 & 0.901 & 0.909 & 0.896 & 0.893\\
    MISSL & 0.317 & 0.252 & 0.398 & 0.277 & 0.254 & \underline{0.315} & \underline{0.241} & \underline{0.399} & \underline{0.267} & \underline{0.254} & 0.901 & 0.896 & 0.908 & 0.893 & 0.890\\
    END4Rec & 0.285 & 0.198 & 0.301 & 0.209 & 0.231 & 0.277 & 0.199 & 0.372 & 0.241 & 0.220 & 0.857 & 0.749 & 0.868 & 0.852 & 0.794\\
    \hline
    \textbf{M-GPT} & \textbf{0.369} & \textbf{0.291} & \textbf{0.460} & \textbf{0.321} & \textbf{0.294} & \textbf{0.338} & \textbf{0.259} & \textbf{0.434} & \textbf{0.290} & \textbf{0.263} & \textbf{0.928} & \textbf{0.906} & \textbf{0.941} & \textbf{0.910} & \textbf{0.902}\\
    Impr. & 12.2\% & 11.1\% & 12.2\% & 12.2\% & 11.4\% & 7.3\% & 7.5\% & 8.8\% & 8.6\% & 3.5\% & 2.2\% & 0.8\% & 3.2\% & 1.2\% & 0.7\%\\
  \bottomrule
\end{tabular}
\end{table*}
\paragraph{4.1.3\quad Baselines.}
To comprehensively demonstrate our proposed M-GPT model, we compare our M-GPT with various recommendation baselines. The details of baselines are shown in Appendix \ref{baseline}.
\paragraph{4.1.4\quad Parameter settings}
We implement our proposed model M-GPT using Pytorch. We search the number of interaction-level multi-behavior dependency from {1,2,3,4}. For comparing equity, we refine the parameter setting of each model to get the best performance. We set the max sequence length $N$ to 200 for all of the models. For M-GPT, We select the number of divided sessions $t_1, t_2$ from \{[2,10], [2,20], [4,10], [4,20]\} and the number of preference granularity $[q_{m1},q_{m2}]$ from \{[20,2],[20,4],[10,2],[10,4]\}. We set hyperparameters $\theta_1$, $\theta_2$ as 1e-5 and learning rate as 0.001. Meanwhile, we set training batch size to 64 for Taobao, Retail and 24 for IJCAI. 
\begin{table}
  \caption{ablation study with key modules}
  \small
  \label{tab:freq}
  \begin{tabular}{c|cc|cc}
    \toprule
    \multirow{2}{*}{Model Variants} & \multicolumn{2}{c|}{Taobao} & \multicolumn{2}{c}{IJCAI} \\
    \cline{2-3} \cline{4-5}
    \quad & HR@5 & NDGC@5 & HR@5 & NDGC@5\\
    \midrule
    \textit{w/o} \textbf{IDE} & 0.344 & 0.266 & 0.313 & 0.239\\
    BERT4Rec-MB & 0.209 & 0.170 & 0.241 & 0.179\\
    \hline
    \textit{w/o} \textbf{interaction-level} & 0.347 & 0.266 & 0.318 & 0.239\\
    \hline
    \textit{w/o} \textbf{item-level} & 0.353 & 0.272 & 0.325 & 0.244\\
    MB-GCN & 0.185 & 0.103 & 0.218 & 0.145\\
    NMTR & 0.125 & 0.082 & 0.109 & 0.076\\
    MB-GMN & 0.192 & 0.108 & 0.235 & 0.161\\
    \hline
    \textit{w/o} \textbf{behavior-level} & 0.358 & 0.279 & 0.330 & 0.250\\
    MB-STR & 0.309 & 0.248 & 0.310 & 0.236\\
    \hline
    \textit{w/o} \textbf{MSPG} & 0.316 & 0.252 & 0.312 & 0.237\\
    \textit{w/o} \textbf{coarse-grained} & 0.327 & 0.261 & 0.307 & 0.240\\
    \textit{w/o} \textbf{fine-grained} & 0.353 & 0.284 & 0.325 & 0.243\\
    \textit{w/o} \textbf{MGMHSA} & 0.340 & 0.263 & 0.318 & 0.239\\
    MBHT & 0.320 & 0.253 & 0.306 & 0.238\\
    \hline
    \textit{w/o} \textbf{Multi-order} & 0.337 & 0.261 & 0.309 & 0.240\\
    \textit{w/o} \textbf{MaxPooling} & 0.301 & 0.244 & 0.296 & 0.224\\
  \bottomrule
\end{tabular}
\end{table}
\subsection{Model Comparison}
We conduct comprehensive comparison experiments among M-GPT and all baselines on Taobao, IJCAI and Retailrocket. We report the result on three datasets in Table1. As illustrated in the table, we can conclude: (1)\textbf{Multi-behavior information promotes recommendation results.} As shown in result table, the model utilizing multi-behavior information generally get a better performance on three datasets than those don't, which prove the benefits of considering multi-behavior information. (2)\textbf{Merging sequential method with graph technology promotes user preference modeling.} MBHT use multi-scale transformer and hypergrah in a parallel style to encode the sequential pattern and multi-behavior dependencies. Meanwhile, our M-GPT merge transformer with graph technology in a sequential style. Both models show their superiority when compared with models using single technology. (3)\textbf{Learning multi-behavior dependency at interaction-level contribute to recommendation.} As shown in table 2, M-GPT outperforms the MB-STR, NMTR, MB-GCN and MB-GMN which learn multi-behavior dependency at behavior-level or item-level.(4)\textbf{M-GPT shows its effectiveness for MBSR problem.} As depicted in result table, M-GPT outperforms all the baselines in terms of all metrics and we summarize the advantage. First, compared with existing multi-behavior recommendation methods, M-GPT learns multi-order interaction-level dependencies which models complex correlations at various orders. Moreover, compared with transformer-based methods, our M-GPT learns the interaction-aware sequential pattern enhanced with capturing multi-grained user preference in different time scales, which further contributes to multifaceted sequential pattern learning.
\begin{figure*}
  \centering
  \subfigure[Taobao]{
  \includegraphics[scale=0.55]{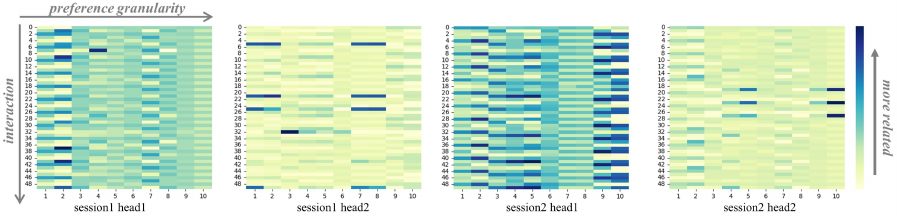}
  }
  \subfigure[IJCAI]{
  \includegraphics[scale=0.55]{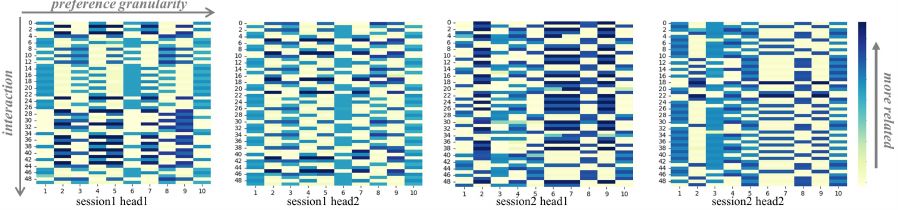}
  }
  \caption{case study on attention map}
  \Description{}
  \vspace{-0.3cm}
\end{figure*}
\begin{figure}
    \centering
    \includegraphics[scale=0.58]{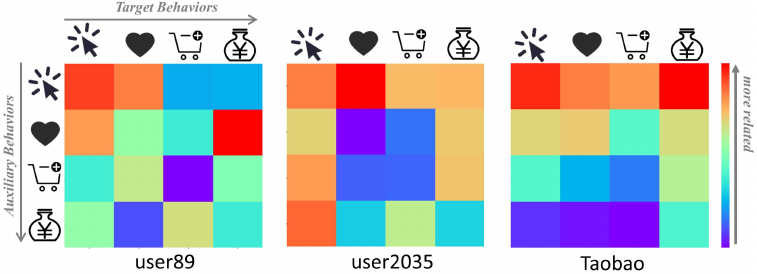}
    \caption{behavioral sequential pattern in Taobao}
    \label{fig:enter-label}
    \vspace{-0.1cm}
\end{figure}
\subsection{Ablation Study}
We have two key modules in our proposed M-GPT including: 1) Interaction-aware Dependency Extractor; 2) Multifaceted Sequential Pattern Generator. Investigating the effectiveness of them is essential for evaluating our model. Meanwhile, we create some variants of M-GPT to further prove the superiority of our design:
\begin{itemize}[labelsep=5pt, leftmargin=10pt]
    \item M-GPT \textit{w/o} IDE: The interaction-aware dependency extractor is replaced by plain item and behavior embedding layer.
    \item M-GPT \textit{w/o} MSPG: This model variant removes multifaceted sequential pattern generator and simply use plain transformer to encode sequential pattern.
    \item M-GPT \textit{w/o} interaction-level: In this variant, the adjacent matrix in IDE is replaced by fully connected adjacent matrix.
    \item M-GPT \textit{w/o} item-level: In this variant, the adjacent matrix is computed by the product of behavior-level representation $B_{i,j}$ and transformation vector $W_i \in \mathbb{R}^d$. 
    \item M-GPT \textit{w/o} behavior-level: In this variant, the adjacent matrix is computed by the product of item-level representation $E_{i,j}$ and transformation vector $W_b \in \mathbb{R}^d$. 
    \item M-GPT \textit{w/o} coarse-grained: In this variant, we just merge global sequential pattern with coarse-grained preference.
    \item M-GPT \textit{w/o} fine-grained: In this variant, we just we just merge global sequential pattern with fine-grained preference.
    \item M-GPT \textit{w/o} MGMHSA: we replace the multi-grained multi-head self-attention with vanilla multi-head self-attention.
    \item M-GPT \textit{w/o} Multi-order: we replace the multi-order multi-behavior dependency by second order output of IDE.
    \item M-GPT \textit{w/o} MaxPooling: We replace the predicting method with attention-weighted sum to corporate the multi-order outputs.
\end{itemize}
We present the results in table3, where we can observe that:(1) \textbf{Each of the two key components contributes to recommendation performance.} As shown in the table 3, there exist a significant performance degradation when M-GPT removes each of its' two key component. (2) \textbf{Ablation study shows the effectiveness of learning multi-behavior dependency at interaction-level.} For IDE, we conduct detailed experiments on three model variants including M-GPT \textit{w/o} interaction-level, M-GPT \textit{w/o} item-level and M-GPT \textit{w/o} behavior-level. The results show the gap of performance between interaction-level dependency learning with other variants. (3) \textbf{Learning multi-grained preference in various time-grained sessions is essential.} For MSPG, we conduct three detailed experiments including M-GPT \textit{w/o} coarse-grained, M-GPT \textit{w/o} fine-grained and M-GPT \textit{w/o} MGMHSA. The results show the effectiveness of proposed MGMHSA in different time granularity. (4) \textbf{Maxpooling is a better method to fuse representations in different orders.} We replace the maxpooling with attention-weighted sum resulting in performance degradation.

\begin{figure}
    \centering
    \includegraphics[scale=0.6]{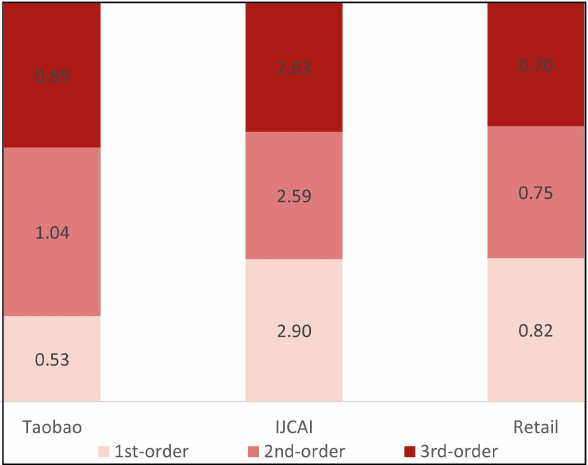}
    \caption{scores in multi-order dependency}
    \label{fig:enter-label}
    \vspace{-0.2cm}
\end{figure}
\subsection{Behavioral Sequential Pattern Analysis}
To investigate the ability of learning users' diverse behavioral sequential pattern, we conduct comprehensive experiment on Taobao dataset. Specifically, we analyze specific sequences from the historical interactions of user89 and user2035. Figure 4 illustrates three 4x4 matrices calculated by the average of attention weights in global sequential pattern. Specifically, we treat behavior pairs and their reversed order (e.g. click-purchase and purchase-click) as different patterns considering various sequential encoding. As shown in figure 5, we can observe the following: 1) Users' behavior patterns vary according to their personal shopping habits. For instance, from figure 5 we can infer that user89 prefers to tag items as favorites before purchasing them. However, it appears that user2035 has no clear preference before making a purchase. 2) In the Taobao dataset, Page-view is the most important behavior. In figure 5 Taobao behavior pattern, Page-view has the highest relevance scores with all the other behaviors, which aligns with the common understanding that Page-view provides the first impression of items to users and triggers them to perform the next behavior.
\subsection{Exploratory Case Study}
To offer an intuitive impression of our model interpretability, we conduct comprehensive case studies on Taobao, IJCAI and Retail dataset. We randomly select 100 items as candidate items for each datasets respectively. First, we study the attention map in case study. As figure 3 depicted, we show the multi-grained attention maps of two attention heads from two sequential sessions in Taobao and IJCAI datasets. It's obvious that the multi-grained preference in each session can be well captured by various query in specific  granularity. At last, we investigate the average scores on 100 candidate items from different multi-behavior dependency orders. We show the results in three datasets in figure 5. In conclusion, we find that learning multi-behavior dependency in different orders truly influences the recommendation performance. Specifically, for Taobao dataset, it's more efficient to capture multi-behavior dependency at 2nd-order. Nevertheless, it seems that there is no notable gap in model performance at multi-order multi-behavior dependency for IJCAI and retail dataset. Moreover, we perform exploratory case study on mulit-behavior dependency in Appendix \ref{multi-behavior}.

\section{Conclusion}
In this paper, we propose a novel model M-GPT for multi-behavior sequential recommendation problem. In interaction-level dependency extractor, we leverage graph-based method to learn the correlations among interactions from item-view and behavior-view information. In multifaceted sequential pattern generator, we learn sequential pattern of user interaction sequence by linear self-attention mechanism and extract multi-grained user preference in different time scales to enrich the representation of sequential pattern. At last, we conduct comprehensive experiments on two public datasets verifying the effectiveness of our proposed M-GPT compared with some state-of-the-art methods. In the future, we aim to explore the capability of Graph Foundation Model (GFM) in Recommendation.
\bibliographystyle{ACM-Reference-Format}
\bibliography{reference}









\clearpage
\appendix
\section{The Learning Process of M-GPT}
\label{algorithm}
\begin{algorithm}
    \caption{: The forward propagation flow of M-GPT}
    \renewcommand{\algorithmicrequire}{\textbf{Input:}}
    \renewcommand{\algorithmicensure}{\textbf{Output:}}
    \begin{algorithmic}[1]
        \REQUIRE The item sequence $e$ and corresponding behavior sequence $b$ for user $u$ with mask tokens at position $I$ and with true labels $T$.
        \ENSURE The estimated likelihood of user $u_i$ engaging with ground-truth items $T$ at specific time step positions $I$.
        
        \STATE  \textbf{\textit{Interaction-Level Multi-Behavior Dependency Extraction}};
        \STATE  Construct interaction-level fully connected graph $\mathcal{G}_{s_i}$ and compute the adjacent matrix $\mathcal{A}$:
        
        $\mathcal{A}_{i,j}\leftarrow E_{i,j} \cdot B_{i,j}, E_{i,j}\leftarrow e_i \odot e_j, B_{i,j}\leftarrow b_i \odot b_j$;
        \STATE  Perform graph convolution on constructed graph to get multi-order dependency representations:  
        
        $H^{(l)}\leftarrow LeakyReLU(\tilde{D}^{-\frac{1}{2}}\tilde{\mathcal{A}}\tilde{D}^{-\frac{1}{2}}H^{(l-1)}W)$;
        \STATE  \textbf{\textit{Multifaceted Sequential Pattern Generator}};
        \STATE  Inject sequential information into multifaceted transformer inputs:

        $H^{(l)}\leftarrow [h_0^{(l)} \oplus p_0,...,h_{N-1}^{(l)} \oplus p_{N-1}]$;
        \STATE  Perform linear self-attention layer to generate global sequential pattern according to equation \ref{10}--\ref{12}:

        $H_{Lin}^{(l)}\leftarrow LinSA(H^{(l)})$;
        \STATE  Perform multi-grained multi-head self-attention layer to generate multi-grained preference representation at different time granularity according to equation \ref{13}--\ref{19}:

        $\tilde{S_{t}^{(l)}}\leftarrow MGMHSA(H^{(l)})$;
        \STATE  Perform projection matrix to fuse multifaceted patterns:

        $ \tilde{H^{(l)}}\leftarrow (H_{Lin}^{(l)}\Vert \tilde{S_{t_1}^{(l)}} \Vert \tilde{S_{t_2}^{(l)}})W^d$;
        \STATE  Perform point-wise feed-forward  to inject non-linearity:

        $\tilde{H^{(l)}}^n\leftarrow LayerNorm(FFN(\tilde{H^{(l)}}^{n-1})+\tilde{H^{(l)}}^{n-1})$;
        \STATE  \textbf{\textit{Maxpooling Prediction}};
        \FOR{each $i \in I, t\in T$}
            \FOR{$m \in [1, l]$}
                \STATE Compute the probability of item at $m$ being $v_t$ under $m$-th order:

                $g_{i,t}^{(m)}\leftarrow \tilde{H^{(m)}}_i^{n}e_t$;
            \ENDFOR
            \STATE Perform maxpooling to search best performance:
            
            $\tilde{p_{i,t}}\leftarrow MaxPooling(g_{i,t}^{(1)},...,g_{i,t}^{(l)})$;
        \ENDFOR
        \RETURN $[\tilde{p_{i_1,t_1}},\tilde{p_{i_2,t_2}},...]$;
    \end{algorithmic}
\end{algorithm}
\section{Time Complexity Analysis}
\label{time complexity}
This section conducts the time complexity analysis of our M-GPT. For the interaction-level dependency extractor, the computation of adjacent matrix $\mathcal{A}$ and graph convolution both take $O(Nd^2)$. For multifaceted sequential pattern generator, the deployment of linear self-attention layer reduce the time complexity of global sequential pattern extraction from $O(N^2d)$ to $O(Nd^2)(N>>d)$ and time-aware multi-grained preference encoding take $O(t\times q_m\times \frac{N}{t}\times d)$, where $t$ represents number of sessions and $q_m$ represents preference granularity. Thus, the overall time complexity of M-GPT is $O((l+1)Nd^2+l(Nd^2+(q_{m_1}+q_{m_2})Nd))$, where $l$ represents the dependency order and $N$ is the max length of each sequences. Based on above discussion, we drop the constant factors in the computation of time complexity and get $O(Nd^2+Nd)$ which is comparable to some SOTA models.
\begin{figure*}
  \centering
  \subfigure[Mask Ratio]{
  \includegraphics[scale=0.2]{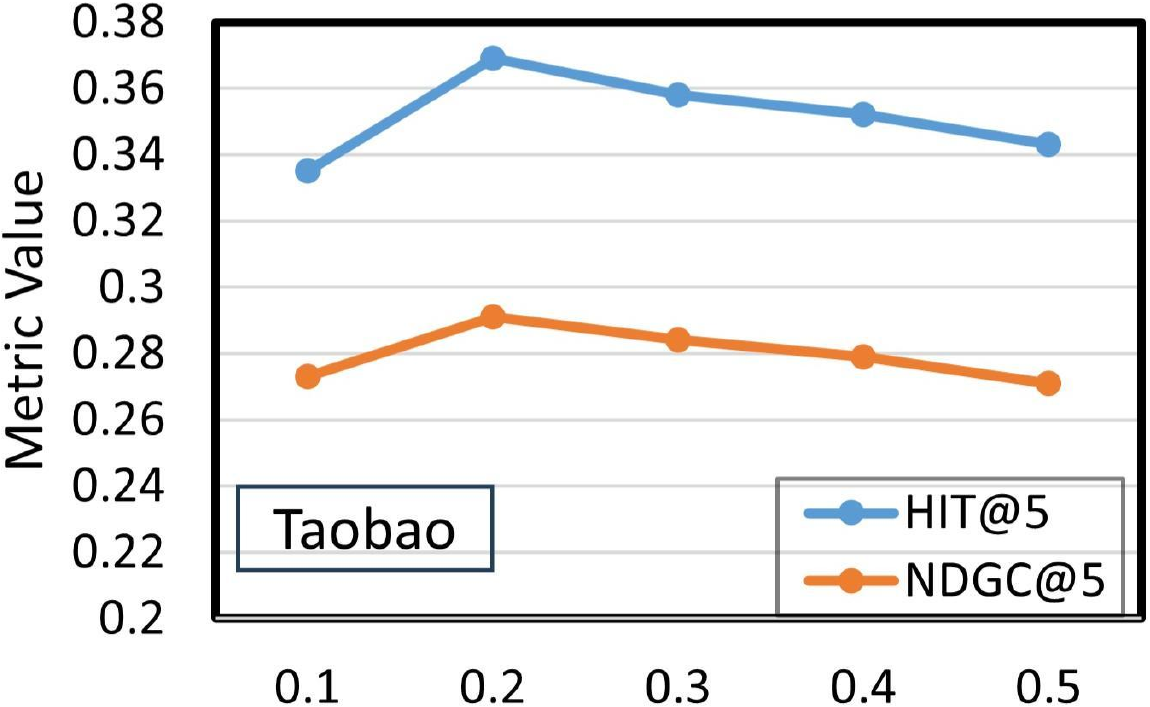}
  }
  \subfigure[Dependency order]{
  \includegraphics[scale=0.2]{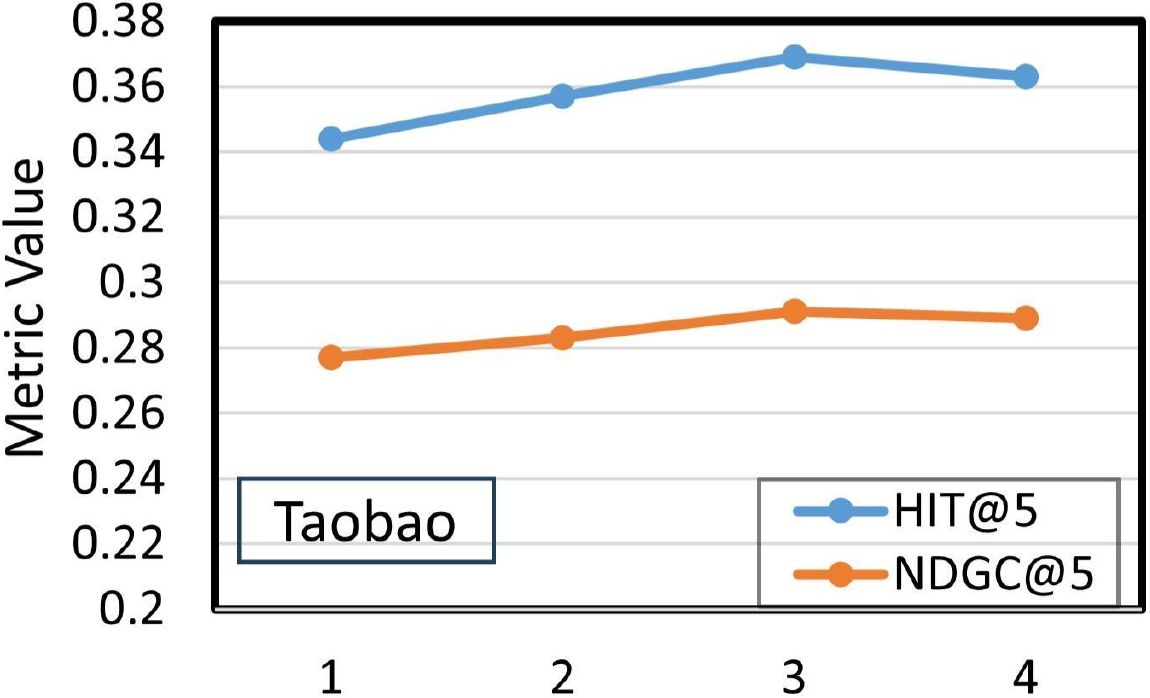}
  }
  \subfigure[Number of Session]{
  \includegraphics[scale=0.2]{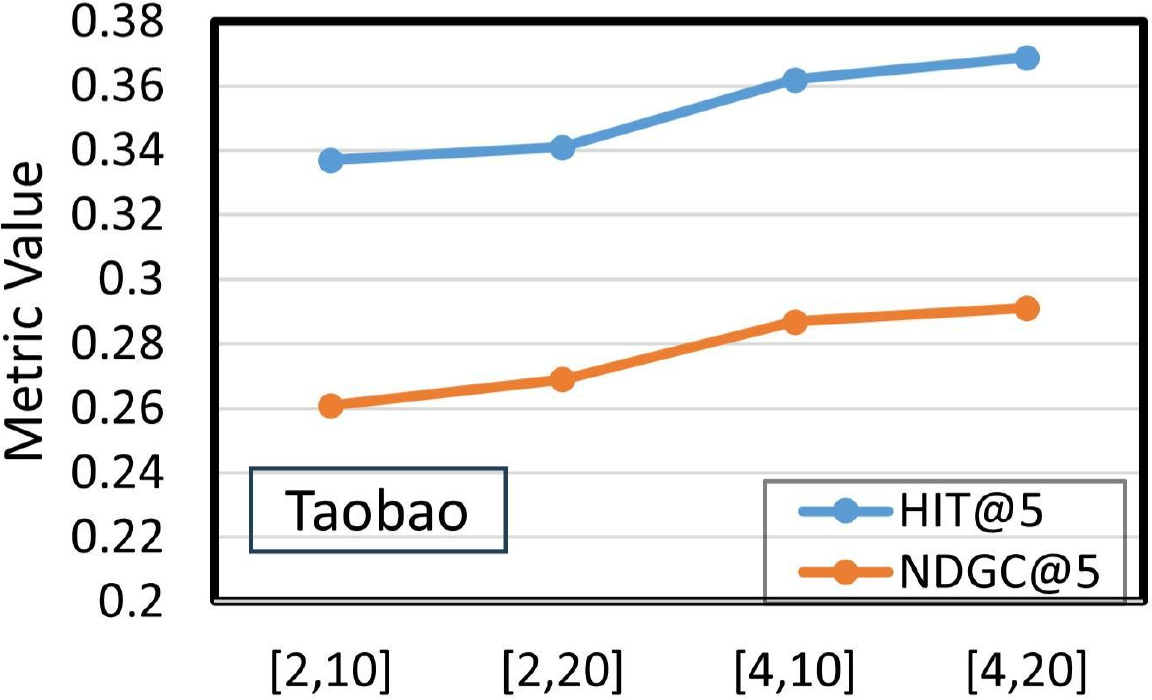}
  }
  \subfigure[Preference Granularity]{
  \includegraphics[scale=0.2]{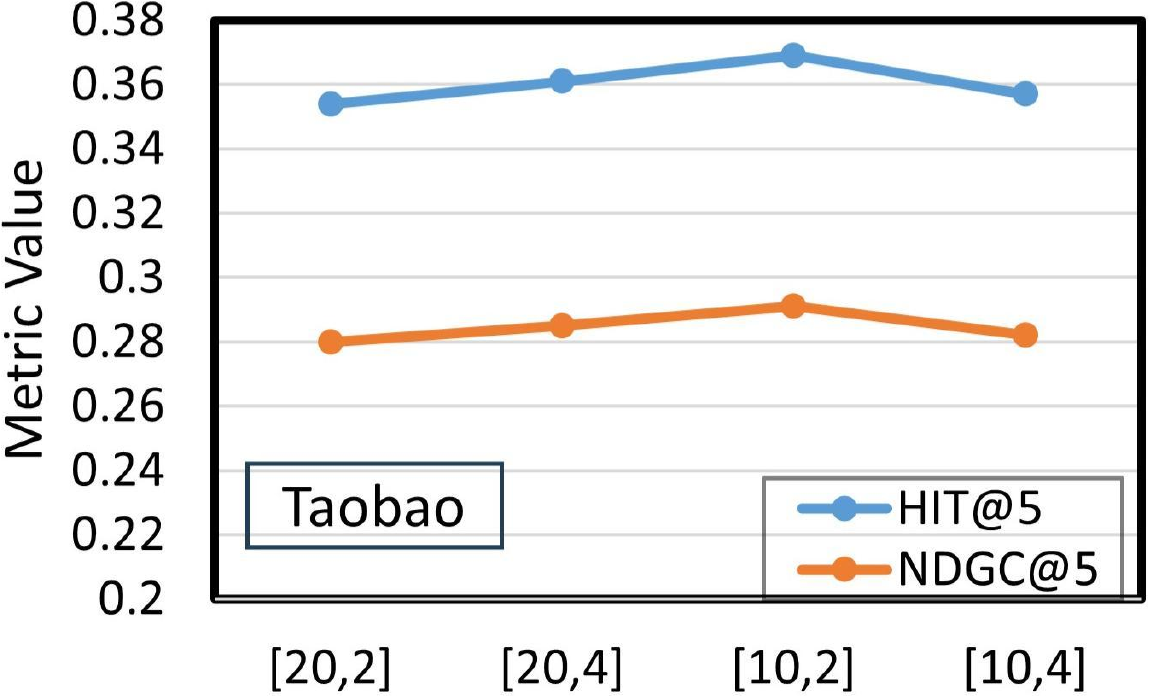}
  }
  \subfigure[Mask Ratio]{
  \includegraphics[scale=0.2]{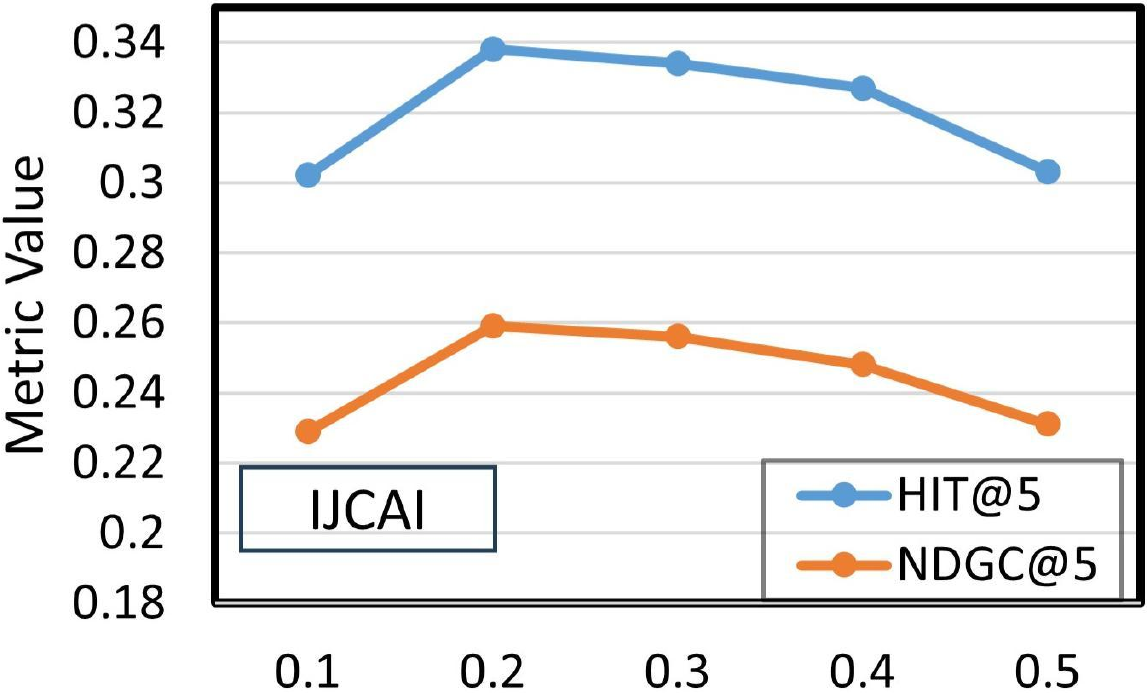}
  }
  \subfigure[Dependency order]{
  \includegraphics[scale=0.2]{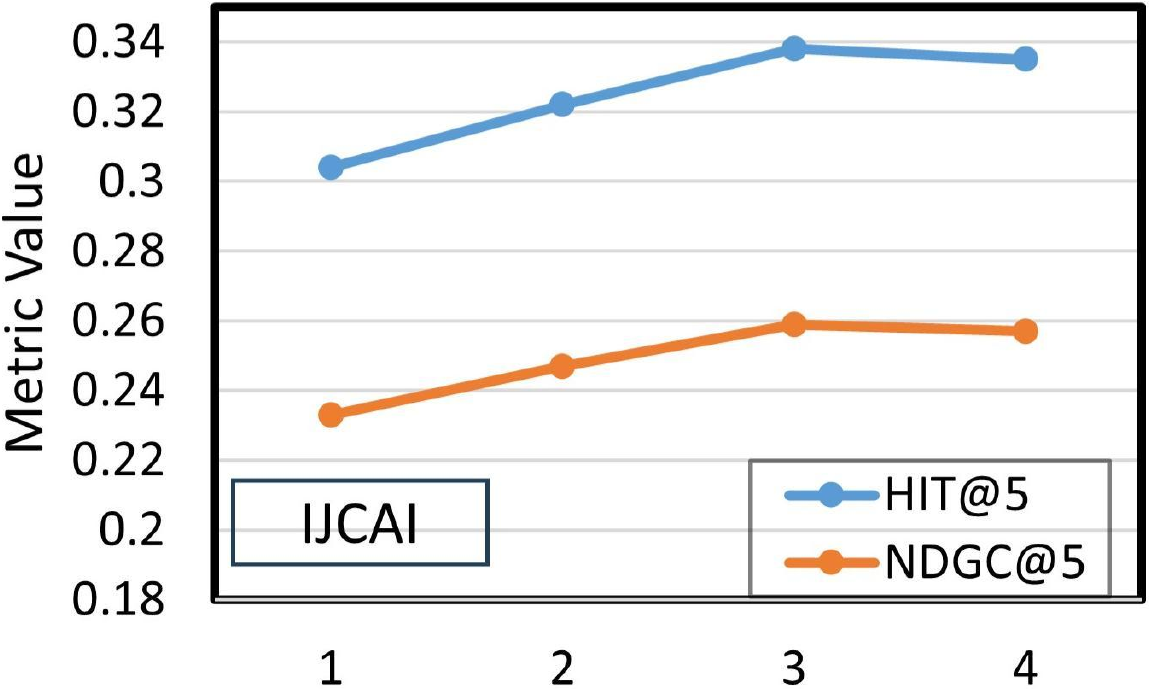}
  }
  \subfigure[Number of Session]{
  \includegraphics[scale=0.2]{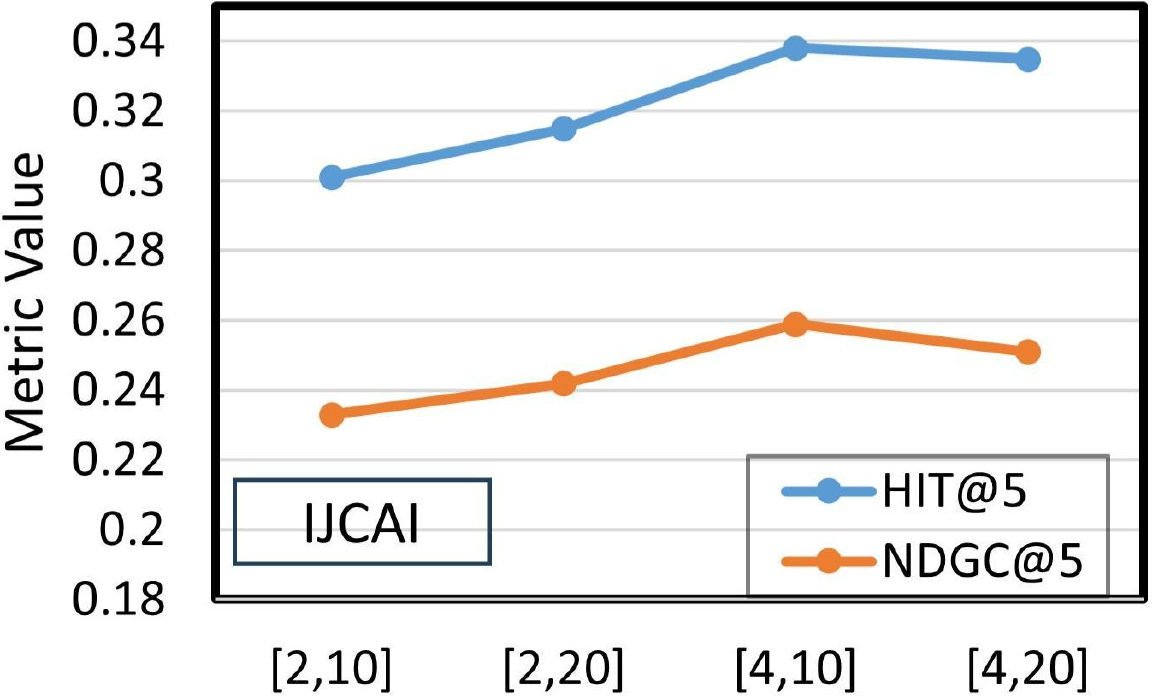}
  }
  \subfigure[Preference Granularity]{
  \includegraphics[scale=0.2]{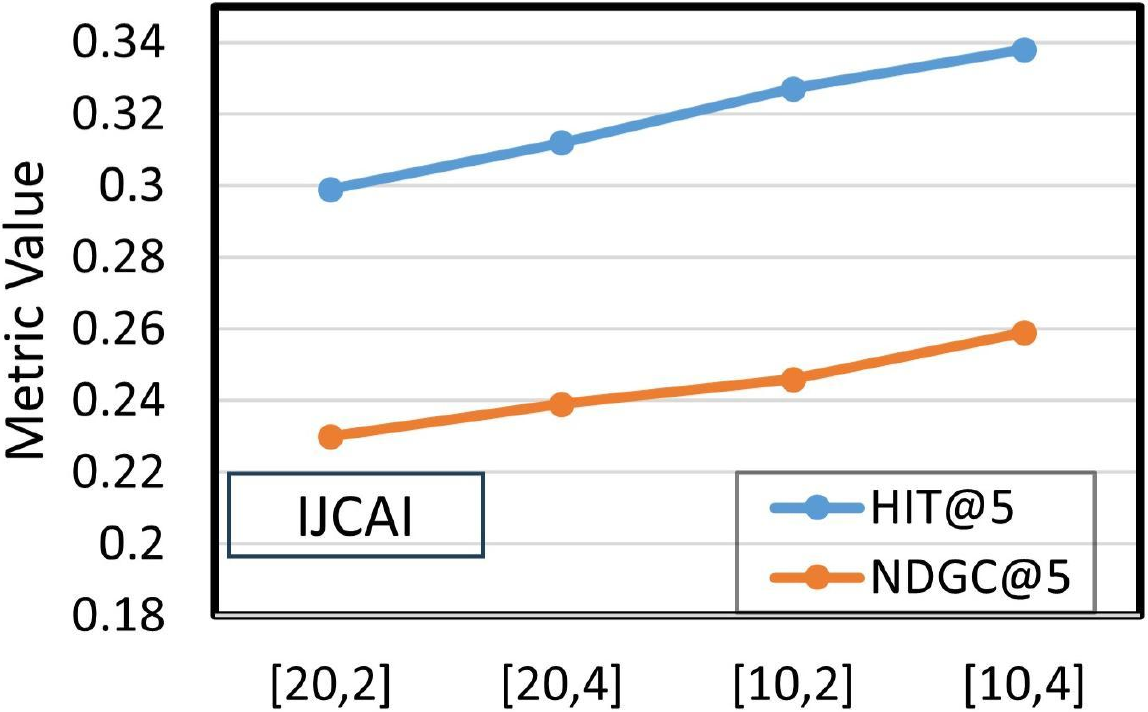}
  }
  \subfigure[Mask Ratio]{
  \includegraphics[scale=0.417]{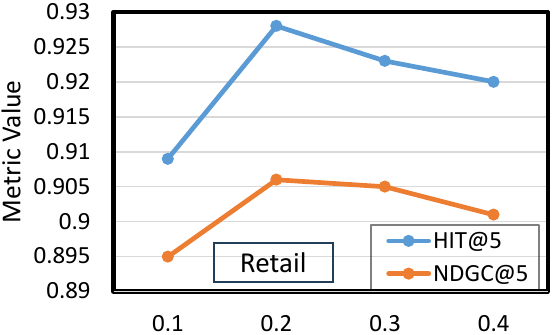}
  }
  \subfigure[Dependency order]{
  \includegraphics[scale=0.417]{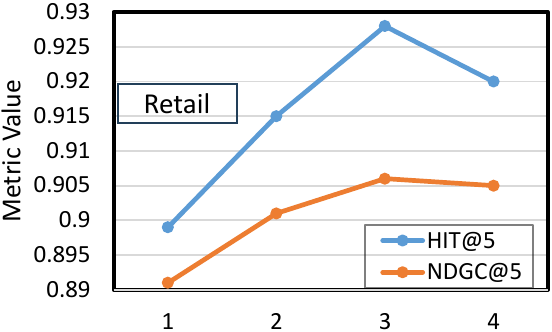}
  }
  \subfigure[Number of Session]{
  \includegraphics[scale=0.417]{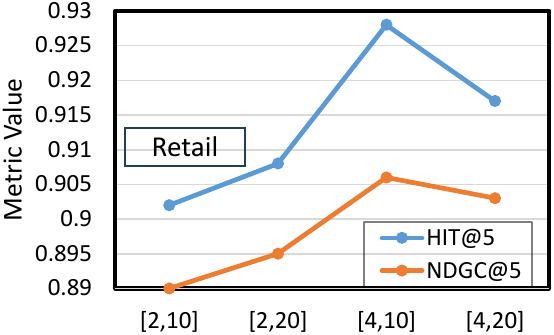}
  }
  \subfigure[Preference Granularity]{
  \includegraphics[scale=0.417]{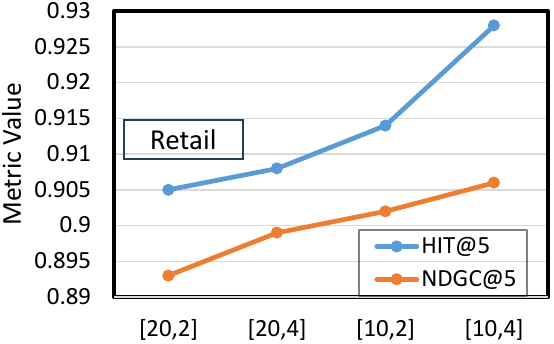}
  }
  \caption{Hyper-parameter analysis on Taobao, IJCAI and RetailRocket}
  \Description{}
  \vspace{-0.1cm}
\end{figure*}

\section{Details of Baselines}
\label{baseline}
\begin{itemize}[labelsep=5pt, leftmargin=10pt]
    \item \textbf{GRU4Rec}\cite{GRU4Rec}. It uses the gated recurrent unit to encode sequential information.
    \item \textbf{SASRec}\cite{SASRec}. It encodes the item-wise sequential relations by self-attention mechanism.
    \item \textbf{Caser}\cite{Caser}. This methods encode the time-evolving user preference by utilizing convolutional neural layers from both vertical and horizontal views.
    \item \textbf{HPMN}\cite{HPMN}. It utilizes a time-evovling hierarchical memory network to model multi-scale transitional information of sequential behaviors.
    \item \textbf{BERT4Rec}\cite{BERT4Rec}. A bidirectional encoder is used for modeling sequential information with Transformer. And the Cloze objective is utilized to optimize the model training.
    \item \textbf{SR-GNN}\cite{SR-GNN}. It generates an item-item graph to perform graph-based message passing to capture local and global user preference.
    \item \textbf{GCSAN}\cite{GCSAN}. It aggregate the self-attention mechanism with GNN structure to better encode graph embedding.
    \item \textbf{HyperRec}\cite{HyperRec}. sequential hypergraphs are used to capture users' dynamic interests.
    \item \textbf{SURGE}\cite{SURGE}. Metric learning is used to construct personalized graphs and hierarchical attention is utilized for extracting multi-dimensional user interets in the graph.
    \item \textbf{BERT4Rec-MB}\cite{BERT4Rec}. We follow the work in MBHT to enhance the BERT4Rec with injecting the behavior type representations into input embedding.
    \item \textbf{MB-GCN}\cite{MBGCN}. Graph convolutional layer is used to enhance the user/item embedding through behavior-aware message passing on the user-item interaction graph.
    \item \textbf{NMTR}\cite{NMTR}. It utilizes a multi-task learning paradigm to model the dependency among different types of behaviors through the behavior-wise cascading relationships.
    \item \textbf{MB-GMN}\cite{MB-GMN}. A graph meta network is used to capture personalized multi-behavior dependency.
    \item \textbf{NextIP}\cite{NextIP}. A Transformer based model that divides the problem into two subtasks: next-item prediction and purchase prediction
    \item \textbf{MB-STR}\cite{MB-STR}. It proposes a multi-behavior transformer framework to model the fine-grained multi-behavior dependency at item level.
    \item \textbf{MBHT}\cite{MBHT}. It proposes a multi-scale transformer enhanced with hypergraphs to capture behavior-aware sequential patterns.
    \item \textbf{DyMuS+}\cite{DyMuS+}. An enhanced version of DyMuS, where the dynamic GRU constructs its internal hidden states as capsules to further capture item-level correlations.
    \item \textbf{TGT}\cite{TGT}. A multi-behavior SRS that captures short-term user interests with a behavior-aware transformer network and long-term user interests via a temporal graph neural network.
    \item \textbf{PBAT}\cite{PBAT}. A transformer-based model that explores personalized multi-behavior patterns and multifaceted time-evolving collaborations for MBSR problem.
    \item \textbf{MISSL}\cite{MISSL}. It proposes a multi-interest self-supervised learning including a behavior-aware multi-interest encoder and intra- and inter-interest self-supervised learning.
    \item \textbf{END4Rec}\cite{END4Rec}. It efficiently captures intricate patterns in user behavior by eliminating noise from user behavior data through noise-decoupling contrastive learning and a guided training strategy.
\end{itemize}

\section{Hyper-parameter Analysis}
\label{hyper}
To evaluate the effectiveness of M-GPT with different settings of hyper-parameters, we conduct a comprehensive experiments on four types of hyper-parameters including \textit{Mask Ratio}, \textit{Number of Dependency Level}, \textit{Number of session} and \textit{Number of Preference Granularity}.
\begin{itemize}[labelsep=5pt, leftmargin=10pt]
    \item \textbf{Mask Ratio} $\rho$. Mask ratio controls proportion of items used as prediction target in sequences. Figure 6a, 6e and 6i show the performance of M-GPT when the mask ratio changes from 0.1 to 0.5 on three datasets. The performance first improves and degrades at last. We can observe that M-GPT achieve best performance on three datasets when the value of mask ratio is 0.2.
    \item \textbf{Dependency Order} $l$. The order of interaction-level dependency representations indicates the complexity of personalized behavior pattern. We change the values of level from 1 to 4 for searching the best set of level value. As illustrated in figure 6b, 6f and 6j, M-GPT reaches the best performance when the value of order is 3 for Taobao, IJCAI and Retail.
    \item \textbf{Number of Session} $[t_1, t_2]$. To enhance the representation of sequential pattern, we intend to learn local multi-grained preference in different time scales. For the consideration of time complexity, we conduct experiments on two datasets for the different time scale setting including [2,10], [2,20], [4,10] and [4,20]. Meanwhile, we fix the number of preference granularity to [10,2]. As shown in figure 6c, 6g and 6k, we observe that the values of HIT@5 and NDGC@5 achieve best performance when the settings are [4,20] for Taobao and [4,10] for IJCAI, retail.
    \item \textbf{Preference Granularity} $[q_{m1},q_{m2}]$.To better model users' local preference, we leverage multi-grained attention mechanism by aggregating query representations. We choose the value of preference granularity for each time scale from [20,2], [20,4], [10,2] and [10,4]. Then we fix the value of session as [4,10] and conduct experiments on three datasets. We observe that the best settings of preference granularity are [10,2] for Taobao and [10,4] for IJCAI, retail.
\end{itemize}
In conclusion, after conducting comprehensive experiments on Taobao, IJCAI and RetailRocket, we find the best setting of hyper-parameter to make M-GPT reach its' best performance. Specifically, we set mask ratio $\rho$ as 0.2, dependency level $l$ as 3, number of session $[t_1, t_2]$ as [4,20] for Taobao and [4,10] for IJCAI, Retail and preference granularity $ [q_{m_1}, q_{m_2}]$ as [10,2] for Taobao and [10,4] for IJCAI, Retail respectively. The difference from the setting of number of sessions and preference granularity may be induced by the sparsity of interaction sequence in three different datasets. 
\begin{figure*}
  \centering
  \subfigure[MBHT, IJCAI]{
  \includegraphics[scale=0.35]{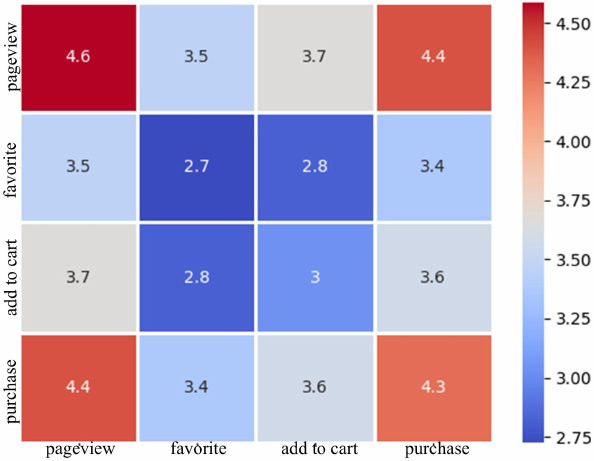}
  }
  \subfigure[M-GPT, IJCAI]{
  \includegraphics[scale=0.35]{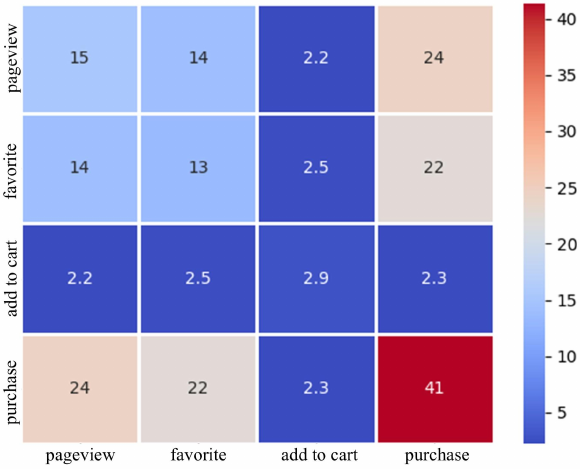}
  }
  \subfigure[MBHT, Taobao]{
  \includegraphics[scale=0.35]{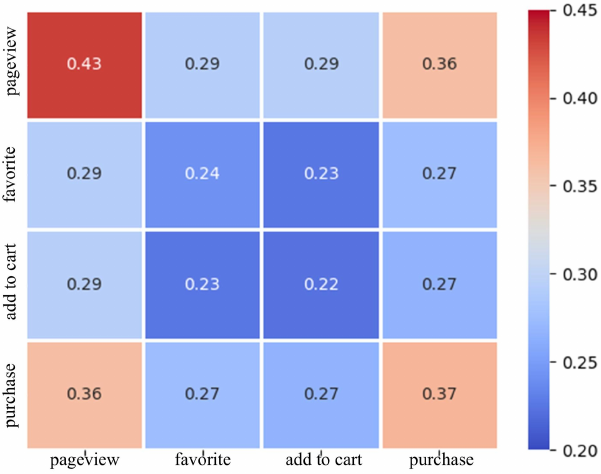}
  }
  \subfigure[M-GPT, Taobao]{
  \includegraphics[scale=0.35]{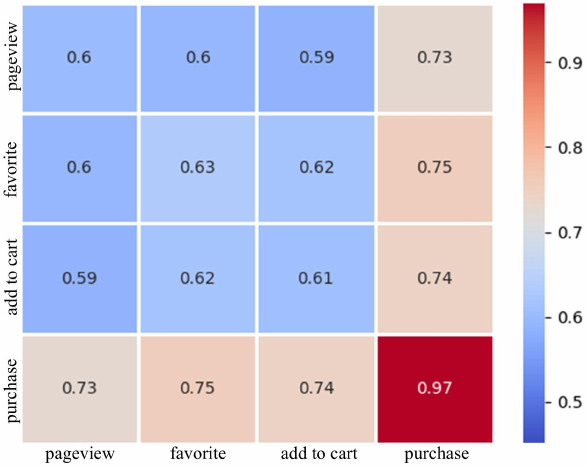}
  }
  \caption{study on behavioral relationship}
  \Description{}
  \vspace{-0.3cm}
\end{figure*}

\begin{figure*}
  \centering
  \subfigure[incidence matrix of user1]{
  \includegraphics[scale=0.68]{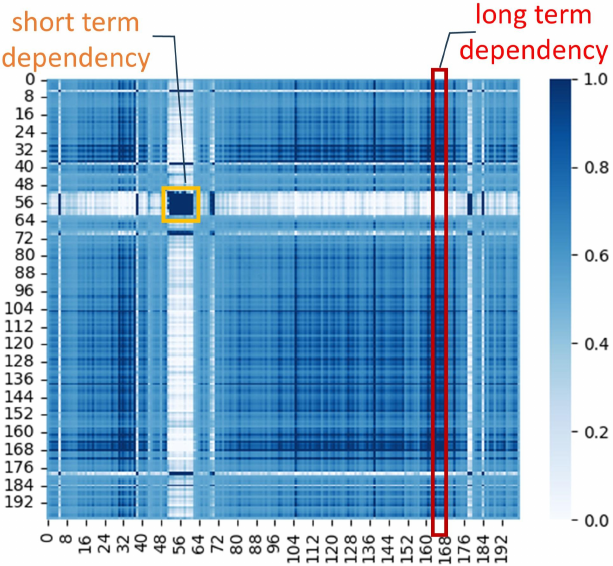}
  }
  \subfigure[incidence matrix of user23]{
  \includegraphics[scale=0.68]{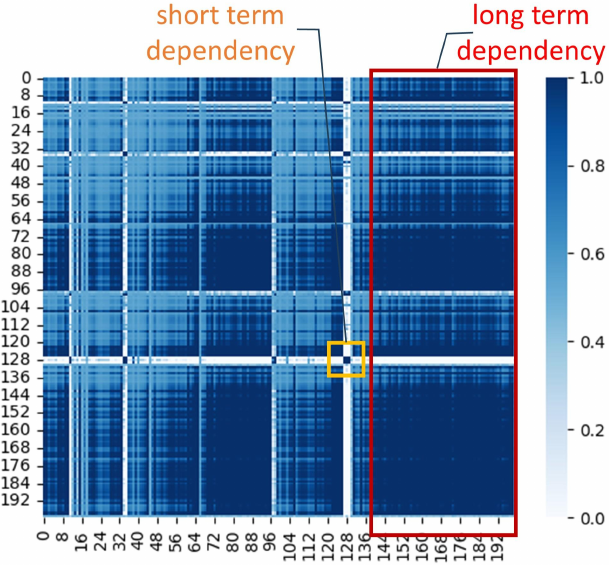}
  }
  \caption{case study on interaction-level multi-behavior dependency}
  \Description{}
  \vspace{-0.3cm}
\end{figure*}
\section{MULTI-BEHAVIOR DEPENDENCY ANALYSIS}
\label{multi-behavior}
\subsection{Behavioral Relationship}
We present the $4\times4$ behavioral relationship matrices for MBHT and M-GPT across two datasets, calculated using the $BB^T$ method. Based on these results, several key observations can be made: 1) The relational values between each pair of behaviors in M-GPT are significantly higher than those observed in MBHT. This underscores M-GPT's advantage in explicitly modeling dependencies at the behavior level, capturing multi-behavior interactions more effectively. 2) Specifically, the relational values for 'favorite-purchase' and 'add to cart-purchase' pairs in M-GPT are markedly higher compared to MBHT. Given that users are generally more inclined to purchase items they have favorited or added to their carts, these findings provide insight into the superior performance of M-GPT in reflecting real-world user behavior patterns. 3) In the IJCAI dataset, the relational scores for 'add to cart' interactions are notably low. This phenomenon could be attributed to the relatively sparse occurrence of 'add to cart' events within this dataset, suggesting that the frequency of specific behaviors may influence the strength of their detected relationships.

In summary, the analysis of the behavioral relationship matrices highlights M-GPT's capability to better capture complex interdependencies among user behaviors. Furthermore, the lower relational scores for certain behaviors, such as 'add to cart', emphasize the importance of considering behavior frequency when interpreting model outcomes.

\subsection{Interaction-Level Dependency}
We also present the incidence matrices of our interaction-level graph in Figure 8. These matrices encapsulate the values of interaction-level dependencies within users' historical sequences, calculated using equations (3)–(5). Several interesting observations can be drawn from these figures: 1) Both users exhibit strong long-term dependency interactions over time. 2) Short-term strong interaction-level dependencies are evident in both users' sequences, likely attributable to temporary deviations in user preferences. 3) Notably, User 23 demonstrates a more stable and stronger multi-behavior dependency at the interaction level compared to User 1, reflecting distinct shopping habits between the two users.

These findings highlight the nuanced patterns of interaction-level dependencies, offering insights into the temporal dynamics and stability of user behaviors. The differences observed between the two users underscore the importance of considering individual behavioral characteristics when analyzing interaction data.
\end{document}